# Design and Modeling of Graded Base III-Nitride HBTs for High-Frequency Applications

Md Saklain Morshed[1], Chandan Joishi[1], Ijaas Mohamed[1], Siddharth Rajan[1,2]
[1]Department of Electrical & Computer Engineering, The Ohio State University, Columbus OH 43210, USA
[2]Department of Materials Science & Engineering, The Ohio State University, Columbus OH 43210, USA
Corresponding author email: *morshed.9@osu.edu*

**Abstract—**This work analyzes the high-frequency performance constraints of AlGaN/GaN heterojunction bipolar transistors (HBTs) and proposes a novel band-engineered graded base N-polar AlGaN/GaN HBT and evaluates the RF performances. Unlike conventional p-GaN base designs, we propose a graded base in the N-polar orientation which offers polarization-induced hole accumulation with quasi-electric field-assisted minority carrier transport. The proposed design simultaneously mitigates the longstanding current crowding, Webster effect, and conventional base resistance limitations. Using well-calibrated two-dimensional TCAD simulations, the proposed compositionally graded base HBT projects peak $f_T/f_{MAX}$ of 190/200 GHz with a power density ($V_{BR}I_{MAX}/8 > 20$ W/mm), offering a promising pathway toward simultaneous high-frequency and high-power GaN millimeter-wave electronics.

## I. Introduction

Heterojunction Bipolar Transistors (HBTs) offer superior linearity [1]–[3] and high current drive capability [4]–[6], making them attractive for radio frequency (RF) applications. Although mm-wave RF technologies are currently dominated by GaN High Electron Mobility Transistors (HEMTs) and GaAs/InP HBTs due to their high cutoff frequency ($f_T$) and maximum oscillation frequency ($f_{MAX}$) [7]–[9], both devices exhibit intrinsic limitations in fully utilizing their material capability. GaN HEMTs suffer from premature surface electric field-induced breakdown [10], while GaAs/InP HBTs are constrained by the relatively lower breakdown field [11]. In contrast, AlGaN/GaN HBTs can utilize the material's high breakdown field since the peak field lies inside the bulk semiconductor [12]. In recent years, GaN HBTs have demonstrated high current gain and breakdown voltage [13], [14]. However, their RF performance remains well below expectations, with the highest reported cutoff frequency ($f_T$) and maximum oscillation frequency ($f_{MAX}$) being around 30 GHz and 5 GHz, respectively [15], [16].

Besides device-scaling limitations, the RF performance bottleneck in III-nitride HBTs can be primarily attributed to two longstanding challenges: high base sheet and contact resistance, which result in current crowding and limit high current-density operation, and the Webster effect arising from insufficient hole concentration in the base, resulting in increased base transit delay and degraded RF performance. In conventional AlGaN/GaN HBTs, the high Mg acceptor activation energy (~150–200 meV) [17] results in a low free hole concentration in the base, leading to high base sheet resistance and degraded carrier transport.

In this work, we show that these limitations are not intrinsic to the III-nitride system and that heterostructure and polarization engineering in the base region can enable simultaneous high-frequency and high-power operation. We show through detailed well calibrated two-dimensional TCAD simulations that N-polar AlGaN/GaN HBTs employing a graded p-GaN to p-AlGaN base can overcome the limitations of conventional GaN HBTs and enable excellent performance up to W-band. Unlike conventional p-GaN base designs, the proposed graded base introduces negative polarization charge, resulting in enhanced hole accumulation, while compositional grading provides a quasi-electric field that accelerates minority-carrier transport and suppresses base recombination. Furthermore, the enhanced hole concentration mitigates the Webster effect under high-current injection. These combined advantages can significantly improve RF performance beyond that of current state-of-the-art GaN HBTs/HEMTs, demonstrating the potential of AlGaN/GaN HBTs as a future platform for high-power millimeter-wave applications.

## II. DEVICE STRUCTURE AND MODELING

We first introduce the concept of the graded-base design by comparing a conventional p-GaN base structure and a structure where the base is graded. Initially, a comparison of the energy band diagram and base conductivity is presented to show how a graded base can improve the hole concentration in the base in Fig.1. The simulation framework is calibrated against the record RF GaN HBT reported by Zhang et al. [15], as shown in Fig.2(a)-(b). The simulated small-signal characteristics and $f_T$/$f_{MAX}$ versus current density show good agreement with reported results, validating the accuracy of the TCAD simulation. For calibration purposes, the material parameters modified from the default Silvaco material library are summarized in Table I. Finally, in the results and discussion section, the device characteristics were presented using two-dimensional TCAD simulations.

TABLE I
IMPORTANT MATERIAL PARAMETERS

| **Parameter** | **Value** |
|---|---|
| Donor activation energy, $E_d$ | 0.01 eV [15] |
| Acceptor activation energy, $E_a$ | 0.147 eV [15] |
| Electron mobility, $\mu_n$ | 80 $cm^2/V\cdot s$ [15] |
| Hole mobility, $\mu_p$ | 6 $cm^2/V\cdot s$ [15] |
| Electron lifetime, $\tau_n$ | 25 ps [18] |
| Saturation velocity, $v_{sat}$ | $2 \times 10^7$ cm/s [19] |
| Auger coefficients, $C_n = C_p$ | $1 \times 10^{-30}$ $cm^6/s$ [20] |
| Radiative recombination coefficient, B | $1 \times 10^{-10}$ $cm^3/s$ [21], [22] |

To compare the epitaxial structures, Fig. 1(a) illustrates a conventional Ga polar AlGaN/GaN HBT with its corresponding energy-band diagrams and hole concentration profiles for p-GaN base thicknesses ranging from 10 to 90 nm (Mg: $3\times10^{19}$ $cm^{-3}$). In contrast, Fig. 1(b) illustrates the proposed graded-base HBT, in which the base composition is graded from p-GaN to 10%p-AlGaN along the N-polar direction while maintaining the same Mg doping. This grading direction is critical because it induces a distributed negative polarization charge within the base [23]–[26]. The resulting polarization induced charge enhances hole accumulation and helps mitigate the Webster effect and Current crowding. For the emitter design of a typical HBT with a non-graded base, a 10-nm-thick $Al_{0.1}Ga_{0.9}N$ layer (Si: $3 \times 10^{19}$ $cm^{-3}$) is incorporated at the emitter–base junction to enhance carrier injection efficiency under forward bias. The remaining 90nm of the emitter is compositionally graded toward GaN (Si: 5× $10^{19}$ $cm^{-3}$) to facilitate ohmic n-GaN emitter contact formation. On the other hand, for graded base HBT, since the base is already graded to 10%AlGaN, emitter composition has started from 20% for wide bandgap emitter. In both devices, the collector and sub collector thicknesses are 300nm (UID: 1 × $10^{16}$ $cm^{-3}$), and 100nm (Si: 1 × $10^{18}$ $cm^{-3}$), respectively. Considering scaled RF HBT, an emitter width of 200 nm and base and collector metal widths of 500 nm are considered. The base metal-to-emitter sidewall and collector metal-to-base sidewall spacings are both set to 500 nm.

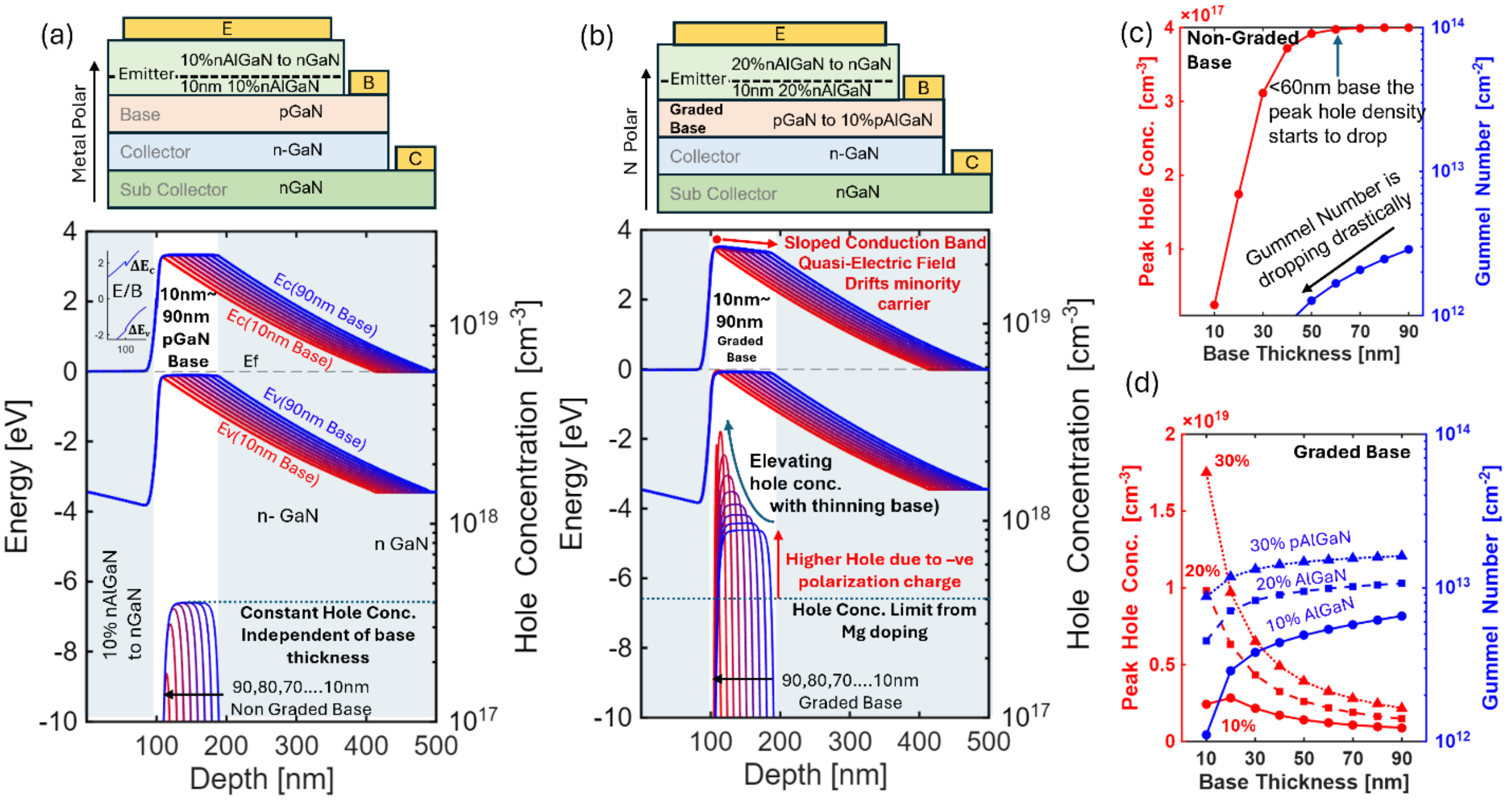

Fig. 1. Epitaxial structures, energy-band diagrams, and corresponding hole concentration profiles of (a) a conventional Ga-polar AlGaN/GaN HBT and (b) the proposed N-polar AlGaN/GaN HBT with a graded p-GaN-to-p-AlGaN base. Dependence of peak hole concentration and gummel number on base thickness for the (c) conventional and (d) proposed HBT bases.

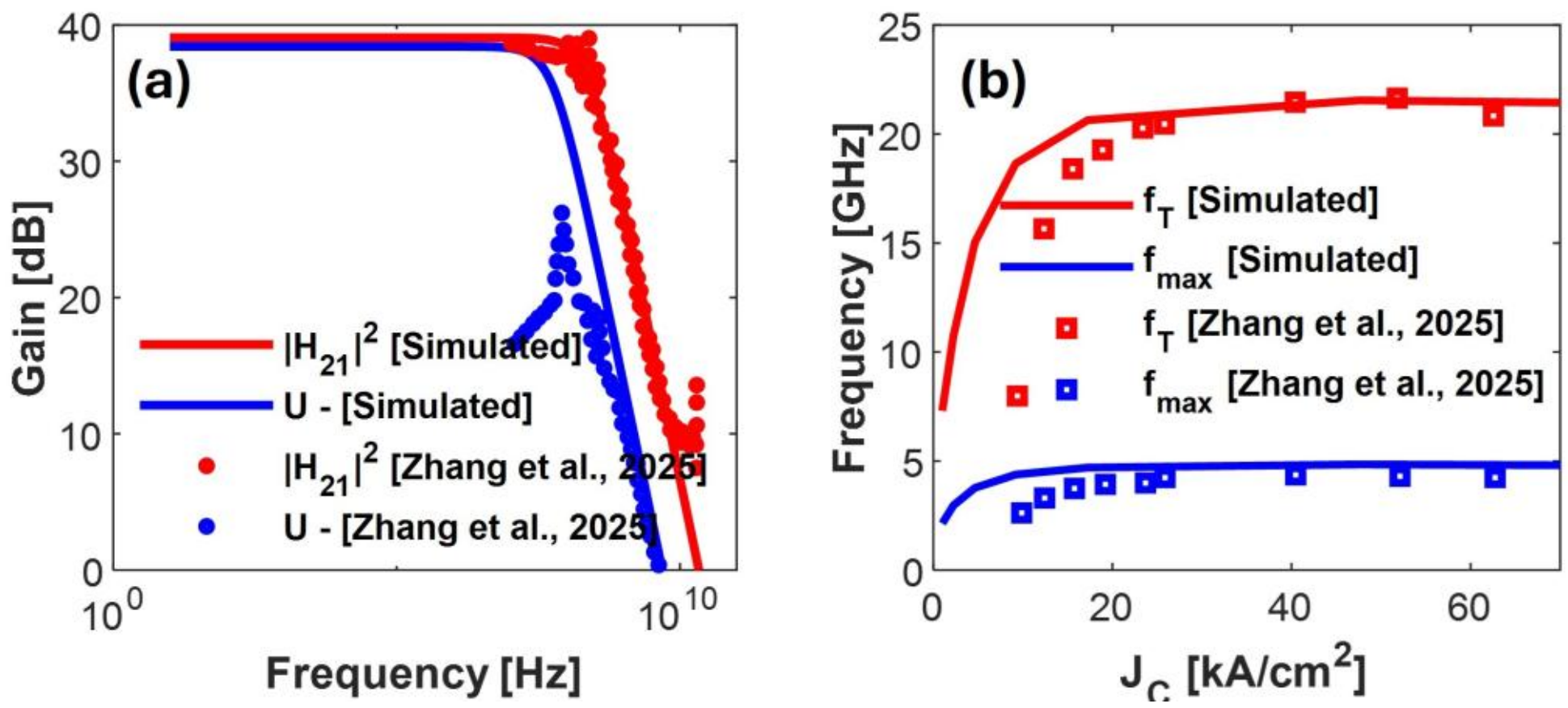

Fig. 2. Calibration of the TCAD simulation framework against the experimental results [15] through (a) the extracted small-signal characteristics and (b) $f_T$ and $f_{MAX}$ as functions of current density.

## III. RESULTS AND DISCUSSION

The RF performance of III-nitride HBTs is primarily determined by the base–emitter and base–collector junction delays, as well as the carrier transit times through the base and collector (Eqn. (1)). Among these contributions, the base transit time is often a dominant limiting factor in underperforming III-nitride HBTs. Since the base transit time scales approximately with the square of the base width ($W_{bn}^2/2D_n$), reducing the base thickness is essential for improving high-frequency performance. However, thinning a conventional p-GaN base reduces the hole sheet density, or Gummel number, thereby increasing the base sheet resistance which does not allow to design a thin base.

$$\tau = \frac{C_{BE}}{g_m} + \frac{C_{BC}}{g_m} + \frac{W_{bn}^2}{2D_n} + \frac{W_c}{2v_{sat}} - (1)$$

Fig. 1(a) and (c) show the dependence of the hole density on the base thickness for typical p-GaN base. Both the volumetric hole concentration and the integrated hole sheet density (Gummel Number) must be considered when evaluating base performance. The volumetric hole concentration influences ambipolar transport and the Webster effect, whereas the sheet charge density/gummel number determines the base conductivity (considering mobility does not change that much), which is crucial for current crowding. Unlike an ideal base, in which the peak hole concentration remains constant before punch-through, the conventional p-GaN base exhibits a sharp reduction in peak hole concentration below 60 nm [Fig. 1(c)]. This reduction is attributed to the relatively long Debye length of holes, which limits charge confinement in thinner bases. Consequently, most reported GaN HBTs employ relatively thick p-GaN bases (70 nm) to maintain sufficient hole concentration and base conductivity [14], [27], [28]. This thick base increases minority carrier recombination as well as base delay significantly. But due to the sharp drop of peak hole density and sheet hole density, the traditional base cannot be shrunk.

In contrast, the graded base exhibits only a slight variation in the integrated hole sheet charge density (or sheet resistance, considering fixed hole mobility) with decreasing base thickness. Because holes are electrostatically induced by polarization rather than relying solely on Mg doping, as shown in Fig. 1(d). Interestingly, the peak hole concentration increases as the base becomes thinner (Fig. 1(d)), opposite to the trend observed in conventional p-GaN bases; this is because the polarization sheet charge density is constant for a fixed AlGaN composition. Consequently, thinning the base increases the volumetric negative polarization charge, which increases the peak hole concentration. This advantage allows the graded base to maintain a nearly constant gummel number over a wide range of base thicknesses, provided that the Al composition remains unchanged. Assuming a hole mobility of 10 $cm^2/V\cdot s$ [29], [30], a 30%-Al graded base exhibits a sheet resistance of approximately 40 kΩ/sq at a thickness of 20 nm, compared with approximately 4000 kΩ/sq for a conventional p-GaN base of the same thickness. This corresponds to an approximately 100-fold reduction in sheet resistance, demonstrating the substantial improvement in base conductivity enabled by polarization-induced hole accumulation for a thin base. In addition to the improved base conductivity, the conduction band gradient establishes a quasi-electric field (Fig. 1(b)) that drifts minority-electron transport toward the collector, thereby reducing the base transit delay [12].

Before discussing the RF characteristics, we first examine the impact of base grading on III-nitride HBT performance through device-physics analysis. We begin by investigating the longstanding current-crowding problem, which leads to localized hotspots under high-current operation and can cause premature device burnout. We then demonstrate how the graded base mitigates current crowding by reducing base resistance.

Furthermore, we analyze how the enhanced hole concentration suppresses minority-electron accumulation, thereby mitigating the Webster effect/Ambipolar transport. Finally, we show how the grading-induced quasi-electric field enhances minority electron velocity through the base, promoting more efficient carrier transport and reducing the base transit time.

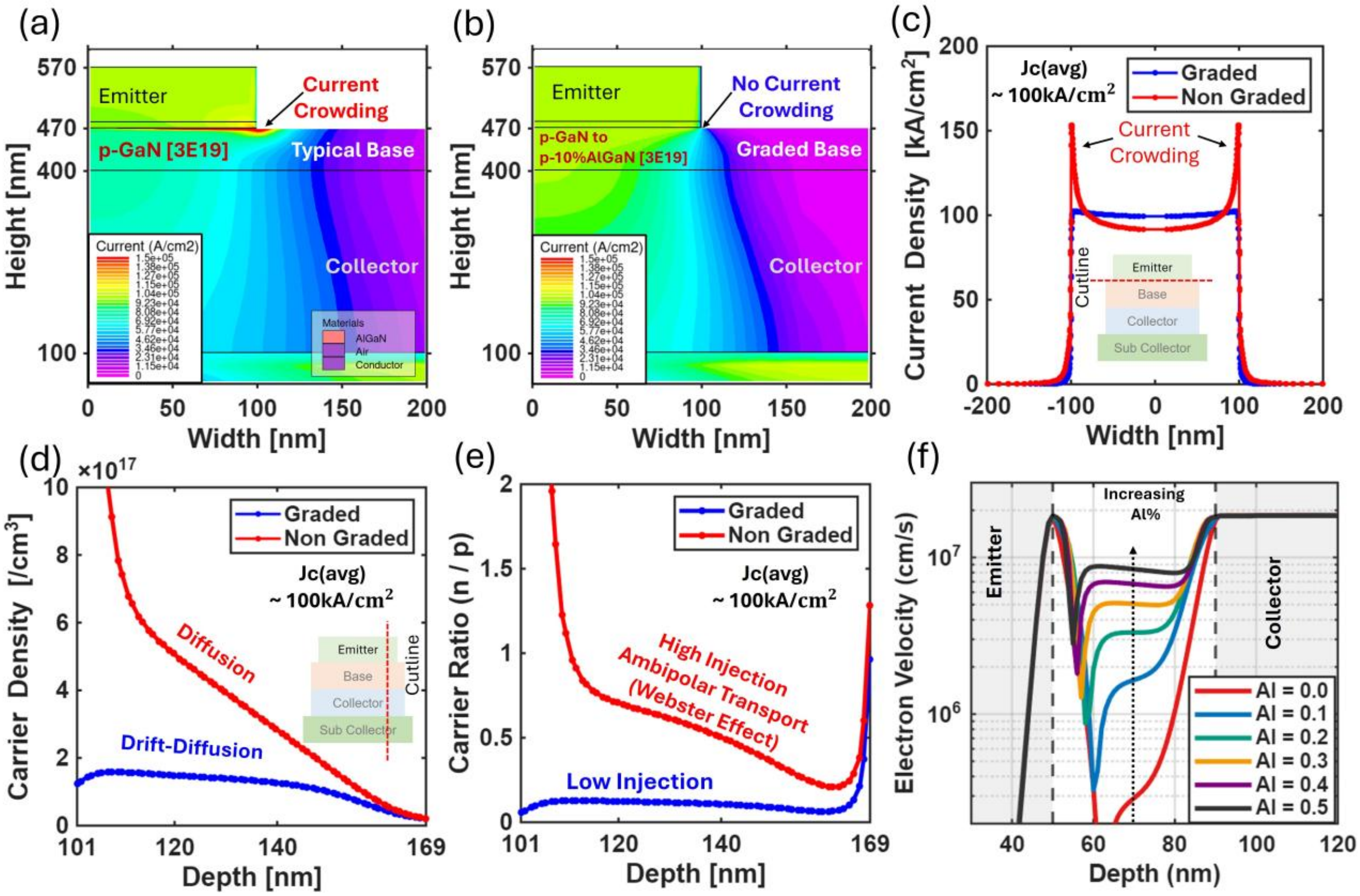

Fig. 3. 2D contour plots of current density at an effective 100 kA/cm$^2$ for (a) 70 nm p-GaN-base HBT and (b) 70 nm p-GaN-to-10% p-AlGaN graded-base HBT; the graded base mitigates current crowding. (c) Current-density cutline beneath the base–emitter junction for both devices. (d) Minority-carrier (electron) concentration at 100 kA/cm$^2$ for conventional and graded bases; the graded base suppresses the minority-electron concentration peak due to quasi-electric-field-assisted drift–diffusion transport. (e) Corresponding electron-to-hole carrier concentration ratio; the graded base maintains a lower-injection profile that avoids ambipolar transport, whereas the conventional base exhibits an ambipolar transport regime. (f) Effect of Al composition on electron velocity in the base; higher Al content strengthens the quasi-electric field, resulting in higher electron velocity and reduced base transit time.

For high-frequency RF applications, HBT must be operated at high current density to obtain high transconductance. This helps to reduce the base–emitter and base–collector delays according to Eqn. (1). However, in typical p-GaN base HBTs, high current density is not common due to early device burnout caused by current crowding. The state-of-the art current density of 275 kA/cm$^2$ HBT is demonstrated by Xia et al. through a novel selective injection method [4]. However, the fundamental problem of low hole density/high sheet resistance has not been solved yet. Fig.3(a)-(c) demonstrates the advantages of the proposed graded-base HBT on forward-active operation at a current density of 100 kA/cm$^2$ compared with a p-GaN base HBT with identical current density. The current-density contours in Fig.3(a) and (b) compare conventional and graded designs with identical 70 nm base, respectively. Owing to the enhanced hole concentration and reduced base sheet resistance (Fig.1(d)), the graded base exhibits a more uniform current distribution, effectively mitigating emitter edge current crowding. This behavior is further confirmed by plotting the current-density profile taking a horizontal cutline beneath the emitter–base junction, where the conventional p-GaN base exhibits severe current crowding with a peak current density approaching 150 kA/cm$^2$ beneath the edge of the emitter (Fig.3(c)). At higher current density operation, this current crowding becomes more severe, creating local hotspots and causing early device failure in traditional AlGaN/GaN HBTs. However, for the graded base, the current density remains almost flat at 100 kA/cm$^2$, which solves the current crowding issue and helps to operate the HBT in high current density operation without any premature burnout.

$$D_a \approx \frac{D_n D_p (n + p)}{D_n n + D_p p} - (2)$$

Another critical consideration is ambipolar transport under high-injection conditions, where the injected minority electron concentration becomes comparable to the majority hole concentration in the base. The low hole diffusion coefficient further limits minority-carrier transport by affecting the ambipolar minority-carrier diffusion coefficient, as shown in Eqn. (2). A quasi-electric field in the graded base flattens the minority electron concentration profile (Fig. 3(d)), reducing charge accumulation and maintaining low-injection operation over a higher current density range. This delays the onset of the Webster effect. As shown in Fig. 3(d)-(e), the graded-base HBT exhibits a substantially lower minority-to-majority carrier concentration ratio, suppressing ambipolar transport (Eqn. (2))

Beyond improving current uniformity and ambipolar transport, high-performance RF HBTs need faster carrier transport in the base. Only diffusion-dominated electron transport in the base is not sufficient to reduce the transit delay. The graded base also enables a quasi-electric field from the gradient in the conduction band that fundamentally helps to transport minority carriers (electrons) faster. As illustrated in Fig. 3(f), the graded base introduces a drift component that increases the carrier velocity in the base. The influence of the quasi-electric field is further quantified through the electron velocity in a 40nm graded base HBT shown in Fig.3(f). Increasing the AlGaN grading composition strengthens the quasi-electric field, resulting in higher electron velocity compared with the diffusion-limited transport in conventional p-GaN base HBTs. This high velocity minority carrier transport reduces the base delay and enables high-frequency performance of N-polar graded-base HBTs.

Finally, after analyzing the device physics using a well calibrated TCAD simulation framework, the impact of enhanced carrier transport is evaluated through RF characteristics. As shown in Fig. 4(a), the conventional p-GaN-base device, with the thinnest base considered (60 nm before punch through, which also exhibits high base resistance), achieves a peak $f_T$ of only 50 GHz, followed by an early roll-off at high current densities due to the onset of the Webster effect. In contrast, the graded-base device (p-GaN to low-composition 5% p-AlGaN, considered to ensure reliable low composition p-AlGaN contact) exhibits significantly improved RF performance, achieving peak $f_T$ values of approximately 85, 105, 140, and 180 GHz for base thicknesses of 60, 50, 40, and 30 nm, respectively shown in Fig. 4(a). The delayed $f_T$ roll-off with decreasing base thickness is attributed to enhanced quasi-electric-field-assisted transport and increased hole density, which suppress the Webster effect and extend high-frequency operation to higher current densities. A similar improvement is observed in the maximum oscillation frequency ($f_{MAX}$). At a base thickness of 30 nm, the peak $f_{MAX}$ reaches 200 GHz, as shown in Fig. 4(b). For the low 5% p-AlGaN composition contact, a standard p-GaN contact resistance of $3.54\times10^{-5}\ \Omega\cdot cm^2$ is considered [52]. Fig. 4(c) shows the extracted $f_T$ as a function of collector current density for different p-GaN-to-p-AlGaN grading compositions. Increasing the AlGaN composition in the graded base further improves $f_T$, above 190 GHz for the p-GaN-to-9% AlGaN base. This enhancement originates from increased polarization-induced hole accumulation and reduced base transit time due to the stronger quasi-electric field, as discussed in Fig. 3(f).

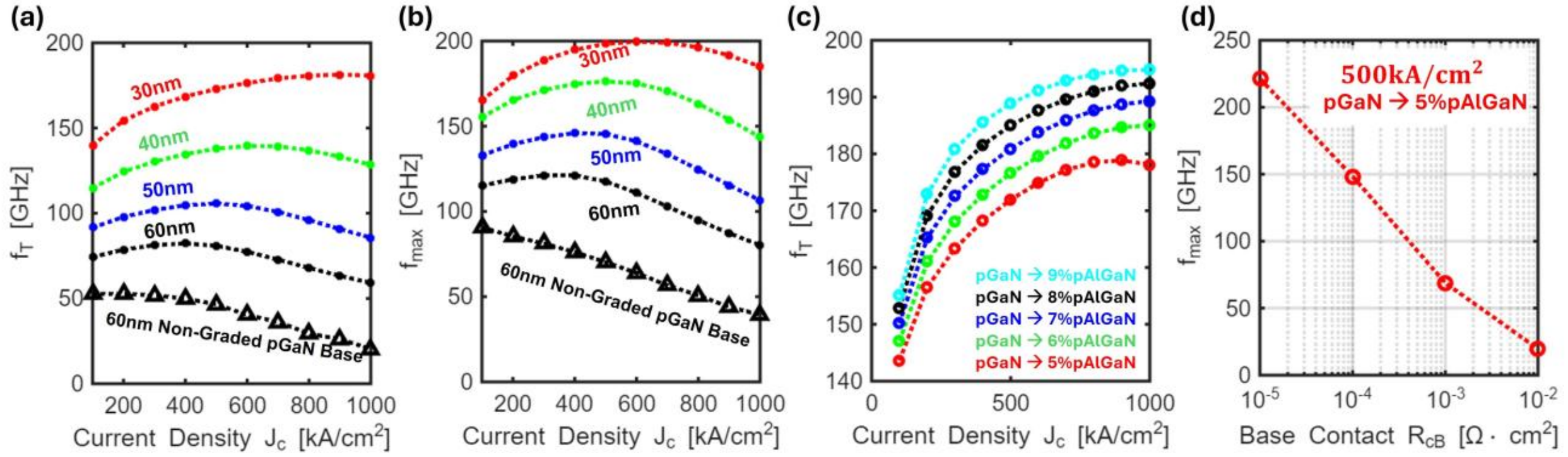

Fig. 4. (a) Simulated (a) $f_T$ and (b) $f_{MAX}$ versus current density for graded-base HBTs with base thicknesses ranging from 60 to 30 nm, compared with the thinnest feasible 60 nm conventional p-GaN-base HBT. (c) Dependence of $f_T$ on AlGaN grading composition. (d) Dependence of $f_{MAX}$ on base specific contact resistance ($\rho_{c,B}$).

Fig. 4(d) evaluates the impact of base contact resistance on $f_{MAX}$ at a fixed current density of 500 kA/cm$^2$. The $f_{MAX}$ degrades with increasing base contact resistance, decreasing from above 200 GHz to below 20 GHz as the specific contact resistance increases from $1\times10^{-5}$ $\Omega\cdot$cm$^2$ to $1\times10^{-2}$ $\Omega\cdot$cm$^2$. This behavior highlights the importance of low-resistance base contacts in maintaining RF performance at high current densities. The combination of polarization-enhanced hole concentration, low base resistance, and optimized contact resistance enables the proposed N-polar graded-base HBT to achieve simultaneous high-frequency and high-power operation.

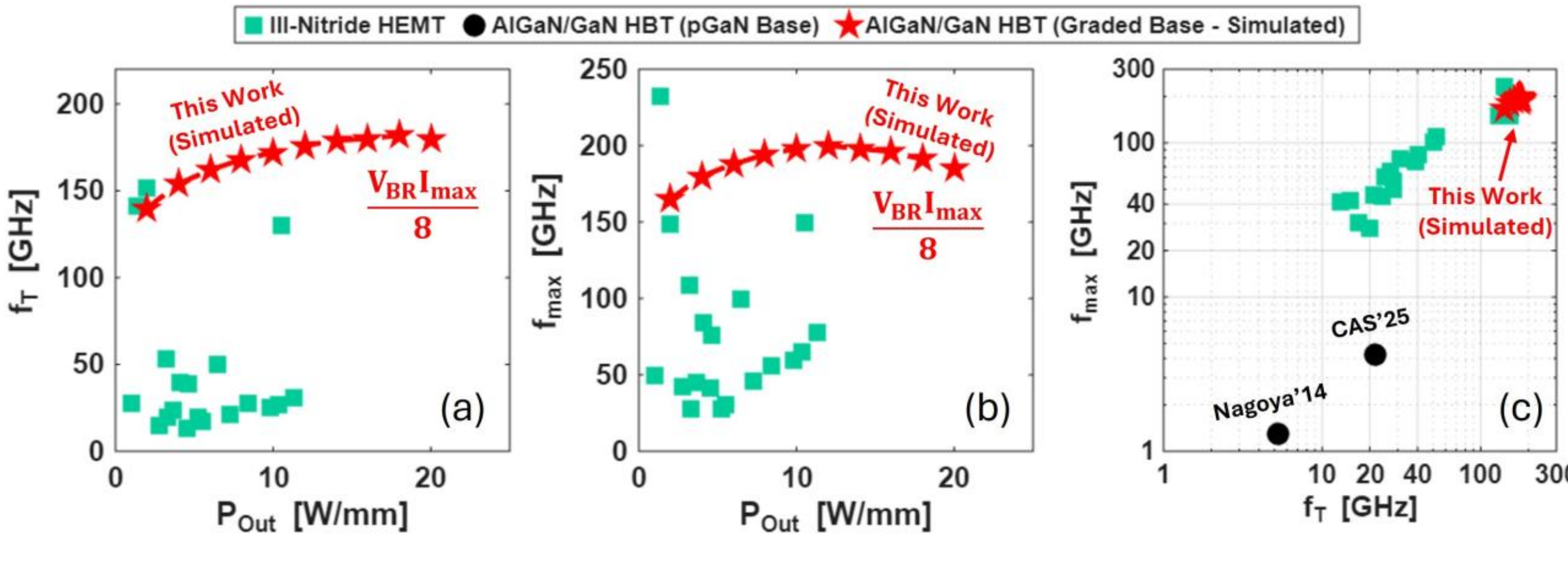

Fig. 5. Benchmark comparison of the proposed AlGaN/GaN HBT with previously reported GaN RF devices: (a) $f_T$ versus output power density (W/mm), (b) $f_{MAX}$ versus output power density (W/mm), and (c) $f_{MAX}$ versus $f_T$ [31]–[51]

To compare the RF performance against state-of-the-art RF devices [31]–[51], benchmark plots are illustrated in Fig. 5. As shown in Fig. 5(a), the proposed graded-base HBT maintains a high cut-off frequency ($f_T$) exceeding 190 GHz while maintaining high output power range. The breakdown voltage has been considered as 80 V for 300nm collector [53]. This demonstrates a significant advantage in high-power-density applications compared to conventional III-nitride HEMTs while maintaining a decent cut-off frequency. Similarly, the maximum oscillation frequency ($f_{MAX}$) benchmarking in Fig. 5(b) confirms that the graded base architecture sustains near 200 GHz, establishing it as a highly competitive candidate for high-frequency power electronics. Finally, the $f_{MAX}$ versus $f_T$ comparison in Fig. 5(c) highlights that while previous AlGaN/GaN HBTs (e.g., Nagoya'14, CAS'25) remained constrained in RF performance, our simulated graded base HBT successfully bridges the performance gap, delivering simultaneous high-speed and high-power capabilities comparable to top-tier III-nitride HEMTs.

## IV. Conclusion

We proposed and investigated the N-polar AlGaN/GaN heterojunction bipolar transistor (HBT) with a linearly graded base, which projects superior RF characteristics compared to conventional HBTs. The enhanced performance arises from the high hole concentration accumulated from induced negative polarization charge, minimized base access resistance, delayed onset of Webster effect and provided quasi-electric field that drifts minority carriers towards collector. These effects collectively mitigate longstanding challenges such as current crowding, high base sheet resistance and Webster effect. For a base thickness of 30nm graded from GaN to 9%AlGaN, the N-polar HBT is predicted to deliver above 20W/mm power density while maintaining a peak $f_T$ of 190GHz and an $f_{MAX}$ of 200GHz.

**Acknowledgements:** We acknowledge funding from AFOSR FA9550-26-1B165 (Program Manager Dr. Michael Yakes) and ARO DEVCOM W911NF2220163 (Program Manager Dr. Tom Oder)

## References

[1] G. Niu, Q. Liang, J. D. Cressler, C. S. Webster, and D. L. Harame, "Rf linearity characteristics of sige hbts," IEEE Transactions on Microwave Theory and Techniques, vol. 49, no. 9, pp. 1558–1565, 2001.

[2] M.-C. Tu, H.-Y. Ueng, and Y.-C. Wang, "Performance of high-reliability and high-linearity ingap/gaas hbt pas for wireless communication," IEEE Transactions on Electron Devices, vol. 57, no. 1, pp. 188–194, 2009.

[3] N. L. Wang, W. J. Ho, and J. Higgins, "Algaas/gaas hbt linearity characteristics," IEEE Transactions on Microwave Theory and Techniques, vol. 42, no. 10, pp. 1845–1850, 1994.

[4] Z. Xia, C. Joishi, S. H. Sohel, A. Xie, M. S. Morshed, B. Liu, J. Hwang, E. Beam, Y. Cao, and S. Rajan, "Selective-injection gan heterojunction bipolar transistors with 275 ka/cm 2 current density," IEEE Electron Device Letters, 2026.

[5] C. Joishi, S. I. Rahman, Z. Xia, S. H. Sohel, and S. Rajan, "Selective injection algan/gan heterojunction bipolar transistors with patterned regrown base contacts," APL Electronic Devices, vol. 1, no. 2, 2025.

[6] M. Dvorak, C. Bolognesi, O. Pitts, and S. Watkins, "300 ghz inp/gaassb/inp double hbts with high current capability and bv/sub ceo/¿ 6 v," IEEE Electron Device Letters, vol. 22, no. 8, pp. 361–363, 2001.

[7] L. Gill, S. DasGupta, J. C. Neely, R. J. Kaplar, and A. J. Michaels, "A review of gan hemt dynamic on-resistance and dynamic stress effects on field distribution," IEEE Transactions on Power Electronics, vol. 39, no. 1, pp. 517–537, 2023.

[8] M. Micovic, D. Brown, D. Regan, J. Wong, Y. Tang, F. Herrault, D. Santos, S. Burnham, J. Tai, E. Prophet et al., "High frequency gan hemts for rf mmic applications," in 2016 IEEE International Electron Devices Meeting (IEDM). IEEE, 2016, pp. 3–3.

[9] M. Urteaga, Z. Griffith, A. Arias-Purdue, A. Carter, P. Rowell, J. Hacker, and B. Brar, "Inp hbt technologies for sub-thz communications," in 2022 IEEE Symposium on VLSI Technology and Circuits (VLSI Technology and Circuits). IEEE, 2022, pp. 122–123.

[10] J. P. Kozak, R. Zhang, M. Porter, Q. Song, J. Liu, B. Wang, R. Wang, W. Saito, and Y. Zhang, “Stability, reliability, and robustness of gan power devices: A review,” IEEE Transactions on Power Electronics, vol. 38, no. 7, pp. 8442–8471, 2023.

[11] B. Markman and M. Schroter, “Indium-phosphide transistors: A review¨ of current state and suitability for commercial¿ 100-ghz wireless communication systems,” IEEE Microwave Magazine, vol. 25, no. 10, pp. 38–53, 2024.

[12] W. Liu, “Handbook of iii-v heterojunction bipolar transistors,” (No Title), 1998.

[13] S. Yan, J. Wang, H. Wang, X. Sun, X. Li, Q. Sun, and H. Amano, “Regrowth-free ex-situ mg-diffused gan-based hbts with high gain of˜ 200,” IEEE Electron Device Letters, 2026.

[14] L. Zhang, X. Wang, J. Zeng, L. Jia, Z. Cheng, Y. Ai, Z. Liu, W. Tan, and Y. Zhang, “Algan/gan heterojunction bipolar transistors with high current gain and low specific on-resistance,” IEEE Transactions on Electron Devices, vol. 69, no. 12, pp. 6633–6636, 2022.

[15] L. Zhang, J. Xu, J. He, Z. Cheng, W. Wang, C. Liu, X. Wang, C. Mi, W. Tan, and Y. Zhang, “Algan/gan heterojunction bipolar transistors with record f/f= 21.6/4.23 ghz,” IEEE Electron Device Letters, vol. 46, no. 6, pp. 912–915, 2025.

[16] L. Zhang, Y. Zhang, J. He, G. Deng, Y. Huang, Y. Liu, Z. Li, Y. Shi, J. Wei, Y. Wei et al., “Experimental demonstration of s-band gallium nitride heterojunction bipolar transistors with p out= 16.5kw/cm 2,” IEEE Electron Device Letters, 2026.

[17] K. Sumida, K. Kataoka, T. Narita, M. Horita, T. Kachi, and J. Suda, “High mg activation with suppressed mg diffusion during ultra-highpressure annealing via sequential nitrogen-ion implantation in mg-ionimplanted gan,” Applied Physics Express, vol. 18, no. 6, p. 066502, 2025.

[18] C. E. Dreyer, A. Alkauskas, J. L. Lyons, J. S. Speck, and C. G. Van de Walle, “Gallium vacancy complexes as a cause of shockley-read-hall recombination in iii-nitride light emitters,” Applied Physics Letters, vol. 108, no. 14, 2016.

[19] K. Park, M. Stroscio, and C. Bayram, “Investigation of electron mobility and saturation velocity limits in gallium nitride using uniaxial dielectric continuum model,” Journal of Applied Physics, vol. 121, no. 24, 2017.

[20] K. T. Delaney, P. Rinke, and C. G. Van de Walle, “Auger recombination rates in nitrides from first principles,” Applied Physics Letters, vol. 94, no. 19, 2009.

[21] Y.-S. Yoo, T.-M. Roh, J.-H. Na, S. J. Son, and Y.-H. Cho, “Simple analysis method for determining internal quantum efficiency and relative recombination ratios in light emitting diodes,” Applied Physics Letters, vol. 102, no. 21, 2013.

[22] S. S. Konoplev, K. A. Bulashevich, and S. Y. Karpov, “From large-size to micro-leds: scaling trends revealed by modeling,” physica status solidi (a), vol. 215, no. 10, p. 1700508, 2018.

[23] A. Nakajima, Y. Sumida, M. H. Dhyani, H. Kawai, and E. S. Narayanan, “High density two-dimensional hole gas induced by negative polarization at gan/algan heterointerface,” Applied physics express, vol. 3, no. 12, p. 121004, 2010.

[24] L. Zhang, K. Ding, J. Yan, J. Wang, Y. Zeng, T. Wei, Y. Li, B. Sun, R. Duan, and J. Li, “Three-dimensional hole gas induced by polarization in (0001)-oriented metal-face iii-nitride structure,” Applied Physics Letters, vol. 97, no. 6, 2010.

[25] W. Hu, P. Qin, W. Song, C. Zhang, R. Wang, L. Zhao, C. Xia, S. Yuan, Y. Yin, and S. Li, “Ultraviolet light-emitting diodes with polarizationdoped p-type layer,” Superlattices and Microstructures, vol. 97, pp. 353– 357, 2016.

[26] S. Li, T. Zhang, J. Wu, Y. Yang, Z. Wang, Z. Wu, Z. Chen, and Y. Jiang, “Polarization induced hole doping in graded alxga1- xn (x= 0.7 1) layer grown by molecular beam epitaxy,” Applied Physics Letters, vol. 102, no. 6, 2013.

[27] L. S. McCarthy, I. P. Smorchkova, H. Xing, P. Kozodoy, P. Fini, J. Limb, D. L. Pulfrey, J. S. Speck, M. J. Rodwell, S. DenBaars et al., “Gan hbt: Toward an rf device,” IEEE Transactions on Electron Devices, vol. 48, no. 3, pp. 543–551, 2001.

[28] L. Zhang, Z. Cheng, J. Zeng, H. Lu, L. Jia, Y. Ai, and Y. Zhang,

"Algan/gan heterojunction bipolar transistor with selective-area grown emitter and improved base contact," IEEE Transactions on Electron Devices, vol. 66, no. 3, pp. 1197–1201, 2019.

[29] C.-C. Chen, T.-C. Huang, Y.-W. Lin, Y.-R. Lin, P.-H. Wu, P.-W. Liou, H.-Y. Hsieh, Y.-Y. Huang, S. Yang, Y.-R. Wu et al., "Hole mobility behavior in al-gradient polarization-induced p-type algan grown on gan template," Applied Physics Letters, vol. 120, no. 2, 2022.

[30] T. Kumabe, S. Kawasaki, H. Watanabe, Y. Honda, and H. Amano, "Hole mobility limiting factors in dopant-free p-type distributed polarizationdoped algan," Applied Physics Letters, vol. 122, no. 25, 2023.

[31] G. Sullivan, J. Higgins, M. Chen, J. Yang, Q. Chen, R. Pierson, and B. McDermott, "High power rf operation of algan/gan hemts grown on insulating silicon carbide substrates," Electronics Letters, vol. 34, no. 9, pp. 922–924, 1998.

[32] A. Raman, S. Dasgupta, S. Rajan, J. S. Speck, and U. K. Mishra, "Algan channel high electron mobility transistors: Device performance and power-switching figure of merit," Japanese Journal of Applied Physics, vol. 47, no. 5R, p. 3359, 2008.

[33] L. Shen, S. Heikman, B. Moran, R. Coffie, N.-Q. Zhang, D. Buttari, I. Smorchkova, S. Keller, S. DenBaars, and a. U. Mishra, "Algan/aln/gan high-power microwave hemt," IEEE Electron Device Letters, vol. 22, no. 10, pp. 457–459, 2001.

[34] T. Palacios, A. Chakraborty, S. Rajan, C. Poblenz, S. Keller, S. DenBaars, J. Speck, and U. Mishra, "High-power algan/gan hemts for kaband applications," IEEE Electron device letters, vol. 26, no. 11, pp. 781–783, 2005.

[35] N. Sarazin, E. Morvan, M. di Forte Poisson, M. Oualli, C. Gaquiere,` O. Jardel, O. Drisse, M. Tordjman, M. Magis, and S. Delage, "Alinn/aln/gan hemt technology on sic with 10-w/mm and 50% pae at 10 ghz," IEEE Electron Device Letters, vol. 31, no. 1, pp. 11–13, 2009.

[36] K. Zhang, Y. Kong, G. Zhu, J. Zhou, X. Yu, C. Kong, Z. Li, and T. Chen, "High-linearity algan/gan finfets for microwave power applications," IEEE Electron Device Letters, vol. 38, no. 5, pp. 615–618, 2017.

[37] D. Marti, S. Tirelli, A. R. Alt, J. Roberts, and C. R. Bolognesi, "150-ghz cutoff frequencies and 2-w/mm output power at 40 ghz in a millimeterwave algan/gan hemt technology on silicon," IEEE electron device letters, vol. 33, no. 10, pp. 1372–1374, 2012.

[38] S. Maroldt, C. Haupt, W. Pletschen, S. Muller, R. Quay, O. Am-¨ bacher, C. Schippel, and F. Schwierz, "Gate-recessed algan/gan based enhancement-mode high electron mobility transistors for high frequency operation," Japanese Journal of Applied Physics, vol. 48, no. 4S, p.04C083, 2009.

[39] D. Marti, S. Tirelli, V. Teppati, L. Lugani, J.-F. Carlin, M. Malinverni, N. Grandjean, and C. R. Bolognesi, "94-ghz large-signal operation of alinn/gan high-electron-mobility transistors on silicon with regrown ohmic contacts," IEEE Electron device letters, vol. 36, no. 1, pp. 17–19, 2014.

[40] N. Nguyen, M. Micovic, W. Wong, P. Hashimoto, L. McCray, P. Janke, and C. Nguyen, "High performance microwave power gan/algan modfets grown by rf-assisted mbe," Electronics Letters, vol. 36, no. 5, p. 1, 2000.

[41] Y.-F. Wu, A. Saxler, M. Moore, R. Smith, S. Sheppard, P. Chavarkar, T. Wisleder, U. Mishra, and P. Parikh, "30-w/mm gan hemts by field plate optimization," IEEE Electron Device Letters, vol. 25, no. 3, pp. 117–119, 2004.

[42] Y.-F. Wu, D. Kapolnek, J. P. Ibbetson, P. Parikh, B. P. Keller, and U. K. Mishra, "Very-high power density algan/gan hemts," IEEE Transactions on Electron Devices, vol. 48, no. 3, pp. 586–590, 2001.

[43] A. Chini, R. Coffie, G. Meneghesso, E. Zanoni, D. Buttari, S. Heikman, S. Keller, and U. Mishra, "2.1 a/mm current density algan/gan hemt," Electronics letters, vol. 39, no. 7, pp. 625–626, 2003.

[44] N. V. C. G. J.-C. D. J. Y. C. F. S. F. N. Hoel, Virginie and J. Massies, "High-power algan/gan hemts on resistive silicon substrate," Electronics letters, vol. 38, no. 14, pp. 750–752, 2002.

[45] X. C. C. L. X. L. Z. H. K. W. X. L. Luo, Weijun and X. Wang, "High performance 1 mm algan/gan hemt based on sic substrate," Frontiers of Electrical and Electronic Engineering in China, vol. 3, no. 1, pp. 120–122, 2008.

[46] N. W. Y. B. P. R. Manfra, M. and D. M. Tennant, "Unpassivated algan/gan hemts with cw power density of 3.2 w/mm at 25 ghz grown by plasma-assisted mbe," Electronics Letters, vol. 39, no. 8, p. 1, 2003.

[47] C. W. G. H. J. W. J. R. C. F. J. L. e. a. Wang, Xiaoliang, “Growth and characterization of 0.8-μm gate length algan/gan hemts on sapphire substrates,” Science in China Series F: Information Sciences, vol. 48, no. 6, pp. 808–814, 2005.

[48] P. S. J. Y. Lee, Cathy and M. A. Khan, “Algan-gan hemts on sic with cw power performance of ¿4 w/mm and 23

[49] K. Makiyama, S. Ozaki, T. Ohki, N. Okamoto, Y. Minoura, Y. Niida, Y. Kamada, K. Joshin, K. Watanabe, and Y. Miyamoto, “Collapse-free high power inalgan/gan-hemt with 3 w/mm at 96 ghz,” in 2015 IEEE International Electron Devices Meeting (IEDM), 2015, pp. 9.1.1–9.1.4.

[50] R. Sun, J. Lai, W. Chen, and B. Zhang, “Gan power integration for high frequency and high efficiency power applications: A review,” IEEE Access, vol. 8, pp. 15529–15542, 2020.

[51] M. M. W. S. W. P. H. L. M. M. P. J. Nguyen, N. X. and C. J. E. L. Nguyen, “High performance microwave power gan/algan modfets grown by rf-assisted mbe,” Electronics Letters, vol. 36, no. 5, p. 1, 2000.

[52] H. Wang, S. Yan, Z. Xu, Y. Lin, P. Yu, J. J. Park, J. E. Dill, H. Wong, Q. Han, X. Li et al., “Enabling thin p-gan ohmic contacts through ultrathin magnesium deposition and brief thermal annealing,” Applied Physics Letters, vol. 129, no. 6, 2026.

[53] K. Nomoto, H. G. Xing, D. Jena, and Y. Cho, “N-polar gan pn junction diodes with low ideality factors,” Applied Physics Express, vol. 15, no. 6, p. 064004, 2022.